\documentstyle[preprint,aps]{revtex}

\draft
\newcommand{\ba}{\begin{eqnarray}}
\newcommand{\ea}{\end{eqnarray}}
\begin{document}

\title{Partial Dynamical SU(3) Symmetry and the Nature of the Lowest K=0 
Collective Excitation in Deformed Nuclei}
\author{ A. Leviatan and I. Sinai}
\address{Racah Institute of Physics, The Hebrew University,
Jerusalem 91904, Israel}
%\date{\today}

\maketitle

\begin{abstract}
We discuss the implications of partial dynamical SU(3) symmetry (PDS) for 
the structure of the lowest K=0$^{+}$ ($K=0_2$) collective 
excitation in deformed nuclei. We consider an interacting boson model 
Hamiltonian whose ground and $\gamma$ bands have good SU(3) symmetry while 
the $K=0_2$ band is mixed. It is shown that the double-phonon components 
in the $K=0_2$ wave function arise from 
SU(3) admixtures which, in turn, can be determined from absolute 
E2 rates connecting the $K=0_2$ and ground bands. 
An explicit expression is derived for these 
admixtures in terms of the ratio of $K=0_2$ and $\gamma$ bandhead energies. 
The SU(3) PDS predictions are compared with existing data and with 
broken-SU(3) calculations for $^{168}$Er.
\end{abstract}
\vspace{12pt}
%\newpage

The nature of the lowest K=$0^{+}$ [K=$0_{2}$] excitation in deformed 
nuclei is still subject to controversy.  
Recently its traditional interpretation as 
a vibration in the $\beta$ degree of freedom 
\cite{bohr69} has been actively discussed and contested [2-5]. 
The preferential decay of some K=$0_{2}$ bands 
in deformed nuclei to the $\gamma$ band rather than to 
the ground ($g$) band have led Casten and von Brentano 
to suggest that these bands should be understood as phonon excitations built 
on top of the $\gamma$ band \cite{casten94}. 
Such decay pattern is consistent with 
calculations \cite{casten94b} in the interacting boson model 
\cite{ibm87} (IBM) 
and in the dynamic deformation model \cite{kumar95}. 
This new interpretation was subsequently questioned and 
challenged. Burke and Sood have claimed that the observed relative 
E2 strengths could arise from rather minor double-$\gamma$-phonon 
admixtures \cite{burke95}. G\"unther {\it et al.} have argued that the 
empirical evidence presented in \cite{casten94} involves higher-spin levels 
which are sensitive to K admixtures, and have shown that band mixing 
calculations can explain the $K=0_2\rightarrow\gamma$ transitions 
without the assumption of double-$\gamma$-phonon character \cite{gunther96}. 
The most relevant information needed to 
resolve the structure of the $K=0_2$ band lies in absolute transition rates. 
An important step in this debate was therefore the measurement of 
lifetimes of the lowest $2^{+}_{K=0_2}$ level \cite{lehmann98} and the 
measurement via Coulomb excitation of B(E2) values connecting 
the $2^{+}_g$ and $2^{+}_{\gamma}$ states with the $0^{+}_{K=0_2}$ level 
in $^{168}$Er \cite{hart98}. 
This nucleus was recently shown to be a good example of SU(3) partial 
dynamical symmetry (PDS), for which 
the ground and $\gamma$ bands have good SU(3) symmetry, while the lowest 
excited  $K=0_{2}$ band is mixed \cite{lev96}. 
The purpose of this work is to study the 
nature of this band under the assumption of SU(3) PDS and to compare the 
predictions with the above mentioned $^{168}$Er data and with broken-SU(3) 
calculations in the IBM framework.

An IBM Hamiltonian with partial SU(3) symmetry has the form \cite{lev96}
\ba
H \;=\; h_{0}P^{\dagger}_{0}P_{0} + h_{2}P^{\dagger}_{2}
\cdot\tilde P_{2} \quad ~. 
\label{hpds}
\ea
Here $s^{\dagger}$ ($d^{\dagger}$) are monopole (quadrupole) bosons whose 
total number is $N$, the dot implies a scalar product and 
$P^{\dagger}_{0} = d^{\dagger}\cdot d^{\dagger} - 2(s^{\dagger})^2$,
$P^{\dagger}_{2,\mu} =  2\,s^{\dagger}d^{\dagger}_{\mu} 
+ \sqrt{7}(d^{\dagger}d^{\dagger})^{(2)}_{\mu}$ are boson-pairs, 
$\tilde P_{2,\mu} = (-1)^{\mu}P_{2,-\mu}$.
For $h_{0}=h_{2}$ the above Hamiltonian 
is an SU(3) scalar related to the Casimir operator of $SU(3)$, while for 
$h_{0}=-5h_{2}$ it is an $SU(3)$ tensor, $(\lambda,\mu)=(2,2)$. Although
$H$ is not an $SU(3)$ scalar, it has a subset of 
solvable states with good $SU(3)$ symmetry.
The solvable eigenstates belong to the ground and 
$\gamma^{k}_{K=2k}$ bands and are simply selected members of the 
Elliott basis \cite{elliott58} with good $SU(3)$ symmetry, 
$(\lambda,\mu)=(2N-4k,2k)K=2k$. States in other bands are mixed. 
The partial $SU(3)$ symmetry of H is converted into partial 
dynamical $SU(3)$ symmetry by adding to it $O(3)$ rotation terms which lead 
to an $L(L+1)$ splitting but do not affect the wave functions.

The Hamiltonian of Eq. (\ref{hpds}) with $h_0=2h_2=0.008$ MeV 
was used in \cite{lev96} to demonstrate the relevance of SU(3) PDS 
to the spectroscopy of $^{168}$Er. 
The resulting SU(3) decomposition of the lowest bands is 
shown in Fig. 1, and compared to the conventional broken-SU(3) calculations 
of Warner Casten and Davidson (WCD) \cite{WCD} where an $O(6)$ term is 
added to an SU(3) Hamiltonian, and to the consistent-Q formalism 
(CQF) \cite{CQF}, 
where the Hamiltonian involves a non-SU(3) quadrupole operator. 
In the WCD and CQF calculations all states are mixed with 
respect to SU(3). In the PDS calculation, states belonging to the ground 
($K=0_1$) and $\gamma$ ($K=2_1$) bands are pure Elliott states 
$\phi_{E}((2N,0)K=0,L)$ and $\phi_{E}((2N-4,2)K=2,L)$ respectively, 
while the $K=0_2$ band is mixed and has the structure
\ba
\vert L,K=0_2\rangle = A_1\tilde\phi_{E}((2N-4,2)\tilde K=0,L)
+ A_2\tilde\phi_{E}((2N-8,4)\tilde K=0,L)\nonumber\\
+ A_3\phi_{E}((2N-6,0)K=0,L) ~.
\label{pdswf}
\ea
Here $\tilde\phi_{E}$ denote states orthogonal to the solvable 
$\gamma^k_{K=2k}$ Elliott's states. 
For $^{168}$Er ($N=16$) the $K=0_2$
band contains $9.6\%$ $(26,0)$ and 
$2.9\%$ $(24,4)$ admixtures into the dominant $(28,2)$ irreducible 
representation (irrep).
Using the geometric analogs of the $SU(3)$ bands \cite{WC82}, 
$(2N-4,2)K=0 \sim \beta$, 
$(2N-8,4)K=0 \sim (\sqrt{2}\beta^2 + \gamma{^2}_{K=0})$, 
$(2N-6,0)K=0 \sim (\beta^2 - \sqrt{2}\gamma^{2}_{K=0})$, the wave function
of Eq. (\ref{pdswf}) can be expressed in terms of the probability amplitudes
for single- and double- phonon $K=0$ excitations
\ba
A_{\beta} &=& A_1 ~,\qquad
A_{\gamma^2} = (A_2 -\sqrt{2}A_3)/\sqrt{3} ~,\qquad
A_{\beta^2} = (\sqrt{2}A_2 + A_3)/\sqrt{3} ~.
\ea
It follows
that in the PDS calculation, the $K=0_2$ band of $^{168}$Er contains 
admixtures of $12.4\%$ $\gamma^{2}_{K=0}$ and $0.1\%$ $\beta^2$ into the 
$\beta$ mode, {\it i.e.} $12.5\%$ double-phonon admixtures into the dominant
single-phonon component.

General properties of the $K=0_2$ band can be studied by examining 
the general SU(3) PDS Hamiltonian of Eq. (\ref{hpds}). 
In Fig. 2 we show the results (filled symbols connected by solid lines) 
of an exact diagonalization ($N=16$) as a function of $h_0/h_2$. 
The empirical value of the ratio of $K=0_2$ and $\gamma$ 
bandhead energies  
$E(0^{+}_{2})/[E(2^{+}_{\gamma})-E(2^{+}_{g})] = 0.8-1.8$, 
in the rare-earth region \cite{casten94,casten94b} constrains the 
parameters of H to be in the range
\ba
0.7 \leq {h_0\over h_2} \leq 2.4 ~.
\label{range}
\ea
In general the $K=0_2$ wave function retains the form as in 
Eq.~(\ref{pdswf}) and, therefore, a 3-band mixing calculation is sufficient 
to describe its structure. To gain more insight into this band mixing,  
we calculate the matrix elements of $H$ (\ref{hpds}) between large-N 
intrinsic states \cite{lev87}
\ba
\vert\beta\rangle &=& b^{\dagger}_{\beta}\vert c;\,N-1\rangle ~, \quad 
\vert\beta^2\rangle = (1/\sqrt{2})(b^{\dagger}_{\beta})^2
\vert c;\,N-2\rangle ~,\quad
\vert\gamma^2_{K=0}\rangle = d^{\dagger}_{2}d^{\dagger}_{-2}
\vert c;\,N-2\rangle ~,
\nonumber\\
\vert c;\,N\rangle &=& (N!)^{-1/2}(b^{\dagger}_{c})^N\vert 0\rangle ~,\quad 
b^{\dagger}_{c} = (1/\sqrt{3})(s^{\dagger} + \sqrt{2}d^{\dagger}_{0}) ~,
\quad 
b^{\dagger}_{\beta} = (1/\sqrt{3})(d^{\dagger}_{0} -\sqrt{2}s^{\dagger}) ~.
\ea
To order $\sqrt{N}$, the symmetric matrix elements ($M_{ij}$) are  
\ba
M_{\beta,\beta} &=& M_{\beta^2,\beta^2}/2 = \epsilon_{\beta}~, \quad
M_{\gamma^2,\gamma^2} = 2\epsilon_{\gamma} ~,
\nonumber\\
M_{\beta,\gamma^2} &=& -\sqrt{2}M_{\beta,\beta^2} = -4(h_0-h_2)\sqrt{N} ~, 
\quad
M_{\gamma^2,\beta^2} = 0 ~,
\nonumber\\
\epsilon_{\beta} &=& 4(2h_0 + h_2)N ~,\quad
\epsilon_{\gamma} = 12h_{2}N ~.
\label{eband}
\ea
Diagonalization of the $3\times 3$ matrix $M_{ij}$ provides a good estimate
both for the bandhead ratio and for the single- and double-phonon 
probabilities $(A_{\beta})^2$, $(A_{\gamma^2})^2$, $(A_{\beta^2})^2$, 
as shown by the dotted lines in Fig. 2.
When the lowest eigenvalue of the matrix $M_{ij}$ is smaller than  
both $2\epsilon_{\beta}$ and $2\epsilon_{\gamma}$, the eigenvalue equation 
simplifies, and we can derive the following expressions for the 
bandhead ratio
\ba
{E(0^{+}_{2})\over E(2^{+}_{\gamma}) - E(2^{+}_{g})} &=& 1+ y 
-{1\over 4N}\,y^2\,{3+y\over 1 - y^2} ~,
\nonumber\\
y = {2\over 3}\Bigl [ \Bigl ({h_0\over h_2}\Bigr ) - 1 \Bigr ] \;\;&=& \;\; 
{\epsilon_{\beta}\over\epsilon_{\gamma}} - 1 ~,
\label{dely}
\ea
and for the mixing amplitudes 
\ba
A_{\beta} &=& {1\over\sqrt{1+\Delta}} ~,\qquad
A_{\gamma^2} = -{1\over \sqrt{2N}}{y\over (1-y)}A_{\beta} ~,\qquad
A_{\beta^2} = {1\over 2\sqrt{N}}{y\over (1+y)}A_{\beta} ~,
\nonumber\\
\Delta &=& {1\over 4N}\,y^2\Bigl [\, {2\over (1-y)^2} + 
{1\over (1+y)^2}\, \Bigr ]~.
\label{approx}
\ea
These expressions are valid for $|y| < 1 -1/\sqrt{2N}$.  
The corresponding results of this approximation are shown in Fig. 2 
as open symbols 
connected by dot-dashed lines.
For $^{168}$Er, ($h_0=2h_2$, $y=2/3$, $N=16$), Eq. (\ref{dely}) yields 
an estimate of $1.62$ for the bandhead ratio as compared with the exact value 
$1.64$. From Eq. (\ref{approx}) we obtain a  
mixing of $11.1\%$ $\gamma^{2}_{K=0}$ and $0.2\%$ $\beta^2$ into the 
$\beta$ mode in good agreement with the exact results mentioned above.
The quantity $y$ in Eq. (\ref{dely}) measures, for large $N$, the extent to 
which the $K=0_2$ band is above ($y>0$) or below ($y<0$) the $\gamma$ band,   
and signals the deviation from SU(3) 
symmetry. In the $SU(3)$ limit $y=0$ 
($h_0=h_2$, $\epsilon_{\beta}=\epsilon_{\gamma}$), 
there is no SU(3) mixing hence no mixing of double-phonon excitations 
into the $K=0_2$ band ($A_{\gamma^2}=A_{\beta^2}=0$ in Eq. (\ref{approx})).
In general, the $SU(3)$ mixing $(1-A_{\beta}^2)$ is $(1/N)$ suppressed, 
but the mixing can be large when $\vert y\vert\rightarrow 1$ 
($h_0/h_2\rightarrow 2.5$), 
corresponding to $\epsilon_{\beta}/\epsilon_{\gamma}\rightarrow 2$. 
The SU(3) breaking and double-phonon admixture is more 
pronounced for $y>0$ ($h_0/h_2 > 1$, 
$\epsilon_{\beta} > \epsilon_{\gamma}$). This can be understood from 
the expression for $\Delta$ in Eq. (\ref{approx}), which is not symmetric 
about $y=0$. Near the SU(3) limit (small $y$), 
$(1-A_{\beta}^2) \sim \Delta\sim (1/4N)y^2\Bigl [\, 3 + 2y\,\Bigr]$, 
which is larger for $y>0$. This implies that the
two-phonon admixtures are expected to be larger when the $K=0_2$ band 
is above the $\gamma$ band. 
As seen from Fig. 2, for most of the relevant range of $h_0/h_2$, 
Eq. (\ref{range}), corresponding to bandhead ratio in the range $0.8-1.65$, 
the double-phonon admixture 
is at most $\sim 15\%$. Only for higher values of the 
bandhead ratio can one obtain larger admixtures and even dominance of the 
$\gamma^{2}_{K=0}$ component in the $K=0_2$ wave function.

An important clue to the structure of $K=0_2$ collective excitations
comes from E2 transitions. The relevant operator is
\ba
T(E2) \; = \; \alpha\, Q^{(2)} + \theta\, \Pi^{(2)}
\label{e2oper}
\ea 
where $Q^{(2)}$ is the quadrupole $SU(3)$ generator and 
$\Pi^{(2)} = (\,d^{\dagger}s + s^{\dagger}\tilde d \,)$
is a (2,2) tensor under $SU(3)$. 
Since the wave functions of the solvable states are known, it is
possible to obtain analytic expressions for the E2 rates
between them \cite{lev96}. If we recall that only the ground
band has the $SU(3)$ component $(\lambda,\mu)=(2N,0)$, 
that $Q^{(2)}$, as a generator, cannot connect different $SU(3)$ irreps 
and that the $\Pi^{(2)}$ term can connect the $(2N,0)$ irrep only with 
the $(2N-4,2)$ irrep, we obtain the
following expressions for B(E2) values of $\gamma\rightarrow g$ and 
$K=0_2\rightarrow g$ transitions
\ba
B(E2;\gamma,L\rightarrow g,L') =
\qquad\qquad\qquad\qquad\qquad\qquad
\qquad\qquad\qquad\qquad\qquad
\nonumber\\
\theta^2\,{\vert\langle\phi_{E}((2N,0)K=0,L')||\Pi^{(2)}||
\phi_{E}((2N-4,2)K=2,L)\rangle\vert^{2}\over (2L+1)}
\nonumber\\
B(E2;K=0_2,L\rightarrow g,L') =
\qquad\qquad\qquad\qquad\qquad
\qquad\qquad\qquad\qquad\qquad\qquad
\nonumber\\
A_{\beta}^2\,\theta^2\,{\vert\langle\phi_{E}((2N,0)K=0,L')||\Pi^{(2)}||
\tilde\phi_{E}((2N-4,2)\tilde K=0,L)\rangle\vert^{2}\over (2L+1)}
~. \qquad
\label{be2}
\ea
Here $\tilde\phi_{E}(\tilde K=0,L)$ is the state orthogonal to the solvable 
Elliott's state $\phi_{E}(K=2,L)$ in the irrep $(2N-4,2)$.
The Elliott states in Eq. (\ref{be2}) can be 
expressed in terms of the Vergados basis \cite{vergados} for which 
the reduced matrix elements of $\Pi^{(2)}$ are known \cite{SU3,ISA}.
The $E2$ parameter $\theta$ in Eq. (\ref{be2}) can be determined from the 
known $2^{+}_{\gamma}\rightarrow 0^{+}_{g}$ E2 rates, and for
$^{168}$Er is found to be $\theta^2=2.175$ W.u. As seen from Eq. (\ref{be2}), 
the B(E2) values for $K=0_2\rightarrow g$ transitions
are proportional to $(A_{\beta})^2$, hence, 
provide a direct way for extracting the amount of SU(3) breaking and 
the admixture of double-phonon excitations in the $K=0_2$ wave function.
In Table 1 we compare the predictions of the PDS and broken-SU(3) 
calculations with the B(E2) values deduced from
a lifetime measurement of the $2^{+}_{K=0_2}$ level in $^{168}$Er 
\cite{lehmann98} (the indicated range for the B(E2) values 
correspond to different assumptions on the feeding of the level) and 
with the B(E2) values connecting the $2^{+}_{g}$ and 
$2^{+}_{\gamma}$ states with the $0^{+}_{K=0_2}$ level, measured in 
Coulomb excitation \cite{hart98}.
It is seen that the PDS 
and WCD calculations agree well with the lifetime measurement, but 
the CQF calculation under-predicts the 
$K=0_2\rightarrow g$ data. This may be due to the fact that the 
CQF parameters are triggered to spectral properties of 
the ground and $\gamma$ bands. 
On the other hand, all calculations show large deviations from the 
quoted B(E2) values measured in Coulomb excitation. It should be noted, 
however, that there are serious discrepancies between the above two 
measurements. First, H\"artelin 
{\it et al.} \cite{hart98}, based on their Coulomb excitation measurement 
and use of generalized Alaga rule, 
predict a value of $0.058\pm 0.007$ (W.u.) for the 
$2^{+}_{K=0_2}\rightarrow 0^{+}_g$ transition, which is marginally within the 
extreme range of the lifetime measurement of 
Lehmann {\it et al.} \cite{lehmann98}. 
The latter refers to an extreme 
and, therefore, highly unlikely feeding scenario. Second, 
the quoted Lehmann \cite{lehmann98} value of $6.2$ W.u. 
(or $3.1$ W.u. assuming $50\%$ E2 multipolarity) for the 
$2^{+}_{K=0_2}\rightarrow 2^{+}_{\gamma}$ transition, 
translates via the Alaga rule to a value of $21.7$ (or $10.85$) W.u. for the 
$0^{+}_{K=0_2}\rightarrow 2^{+}_{\gamma}$ transition. The latter is a factor 
of $7.8$ (or $3.9$) larger than the value $2.8\pm 0.4$ W.u. of 
H\"artelin \cite{hart98}. 
An independent measurement of the lifetime of the $0^{+}_{K=0_2}$ 
in $^{168}$Er is highly desirable to clarify this issue.

To summarize, we have investigated the nature of the lowest collective $K=0$ 
excitation in deformed nuclei under the assumption of SU(3) partial dynamical 
symmetry (PDS). We have presented three types of calculations: 
an exact diagonalization, a 3-band mixing calculation using intrinsic states, 
and an analytic approximation to the latter. In this framework, 
the SU(3) breaking and double-phonon admixture in the 
$K=0_2$ wave function are intertwined. The mixing is of order ($1/N)$ but 
depends critically on the ratio of the $K=0_2$ and $\gamma$ bandhead 
energies. It can be obtained directly from a knowledge of absolute $E2$ 
rates connecting the $K=0_2$ band with the ground band. 
The PDS predictions agree with the 
lifetime measurement of the $2^{+}_{K=0_2}$ level in $^{168}$Er 
\cite{lehmann98} but a noticeable discrepancy remains with respect to the 
B(E2) values measured via Coulomb excitation \cite{hart98}. 
For the $K=0_2$ wave function in $^{168}$Er, we find 
$12.5\%$ of double-phonon admixtures into the dominant single-phonon
component. These findings support the conventional single-phonon 
interpretation for this band with small but significant 
double-$\gamma$-phonon admixture. 

Useful discussions with R.F. Casten on the empirical data of $^{168}$Er 
are acknowledged. This work is supported in part by a grant from the 
Israel Science Foundation. A.L. thanks the Institute for Nuclear Theory at the 
University of Washington for its hospitality and the Department of Energy for 
partial support during the completion of this work.

\begin{table}
\caption[]{Comparison of theoretical and experimental absolute B(E2) 
values [W.u.] for transitions from the $2^{+}_{K=0_2}$ level [8] and  
to the $0^{+}_{K=0_2}$ level [9] in $^{168}$Er.
\normalsize}
\vskip 10pt
\begin{tabular}{lcc|ccc}
\multicolumn{3}{c|}{Exp.} &
\multicolumn{3}{c}{Calc.}\\
Transition & B(E2) & range & PDS & WCD \cite{WCD} & CQF \cite{CQF} \\
\hline
\multicolumn{3}{c|}{Lifetime measurement [8]} &
\multicolumn{3}{c}{ }\\
$2^{+}_{K=0_2}\rightarrow 0^{+}_g$ & 0.4 & 0.06--0.94  &
0.65  & 0.15  & 0.03  \\
$2^{+}_{K=0_2}\rightarrow 2^{+}_g$ & 0.5 & 0.07--1.27  &
1.02  & 0.24  & 0.03  \\
$2^{+}_{K=0_2}\rightarrow 4^{+}_g$ & 2.2 & 0.4--5.1  &
2.27 & 0.50 & 0.10 \\
$2^{+}_{K=0_2}\rightarrow 2^{+}_{\gamma}$ $^{a)}$ & 6.2 (3.1) 
& 1--15 (0.5--7.5) &
4.08 & 4.16 & 4.53 \\
$2^{+}_{K=0_2}\rightarrow 3^{+}_{\gamma}$ $^{a)}$ & 7.2 (3.6) 
& 1--19 (0.5--9.5) &
7.52 & 7.90 & 12.64 \\
\hline
\multicolumn{3}{c|}{Coulomb excitation [9]} &
\multicolumn{3}{c}{ }\\
$2^{+}_g\rightarrow 0^{+}_{K=0_2}$ & $0.08\pm 0.01$  & &
0.79   & 0.18   & 0.03  \\
$2^{+}_{\gamma}\rightarrow 0^{+}_{K=0_2}$ & $0.55\pm 0.08$  & &
3.06 & 3.20 & 5.29 \\
\end{tabular}
{\small $^{a)}$ The two numbers in each entry
correspond to an assumption of pure E2 
and (in parenthesis) 50\% E2 multipolarity.}
\end{table}

\newpage
\begin{figure}
\caption[]{
$SU(3)$ decomposition of wave functions of the ground ($K=0_1$), 
$\gamma$ ($K=2_1$),
\hfill\break
and $K=0_2$ bands of $^{168}$Er ($N=16$) in the SU(3) PDS calculation  
(present work), and broken-SU(3) calculations WCD \cite{WCD} and CQF 
\cite{CQF}.}
\end{figure}
\noindent
\begin{figure}
\caption[]{Properties of the $K=0_2$ band as a function of $h_0/h_2$, 
parameters of the SU(3) PDS Hamiltonian, Eq.~(\ref{hpds}), N=16.
(a) Ratio of $K=0_2$ and $\gamma$ bandhead energies obtained from an 
exact diagonalization (filled circles), 3-band mixing calculation based on 
Eq.~(\ref{eband}) (dotted line) and an approximation based on 
Eqs.~(\ref{dely})$-$(\ref{approx}) 
(open circles connected by a dot-dashed line).
(b)~Probability amplitudes squared, $(A_{\beta})^2$ (circles), 
$(A_{\gamma^2})^2$ (squares), $(A_{\beta^2})^2$ (triangles down) for the 
$K=0_2$ wave function.
Notation for the different curves as in part (a) with corresponding symbols.} 
\end{figure}

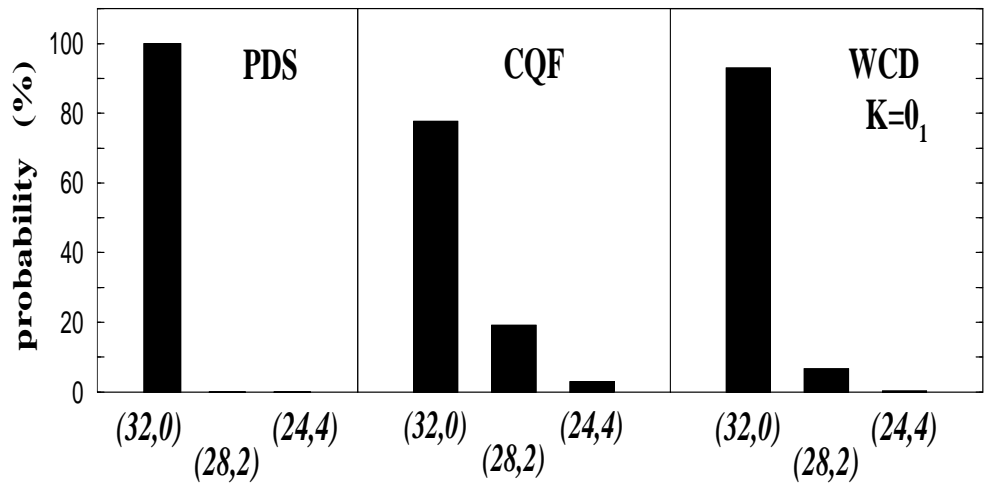
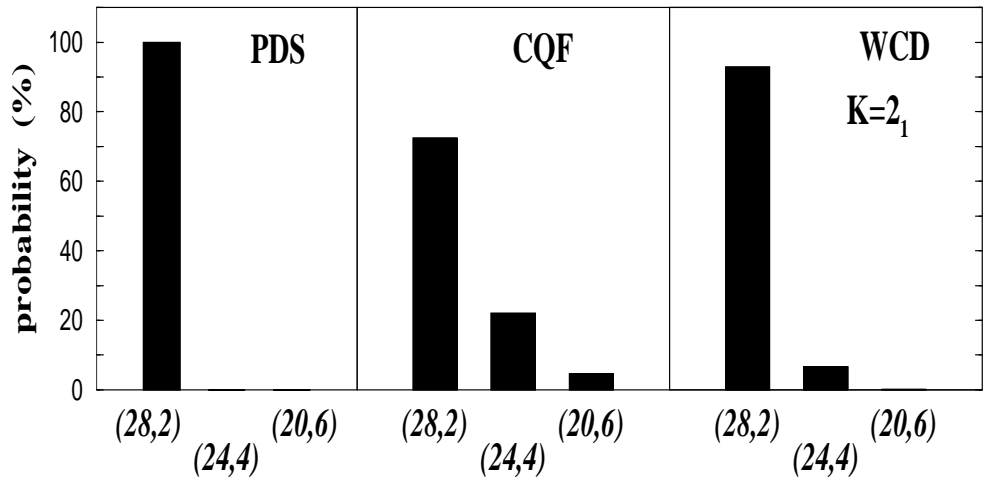
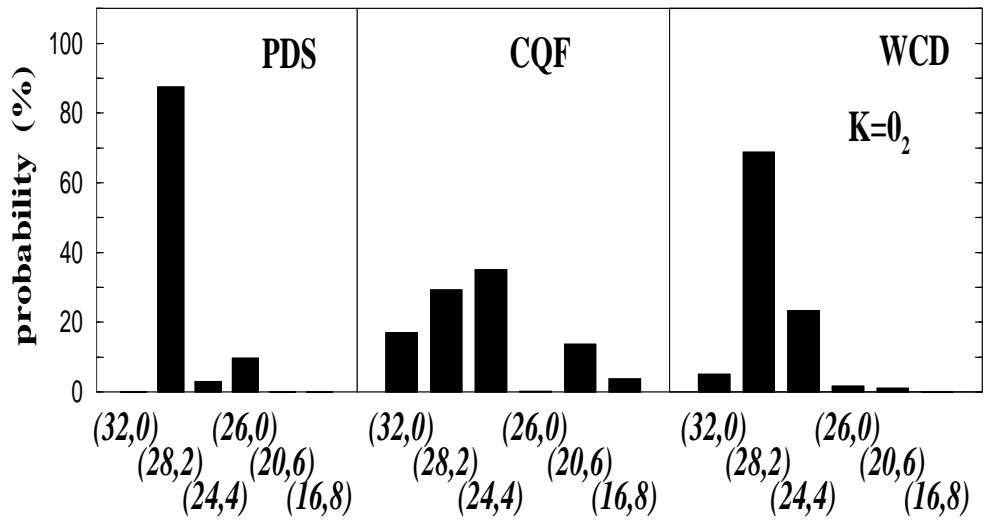
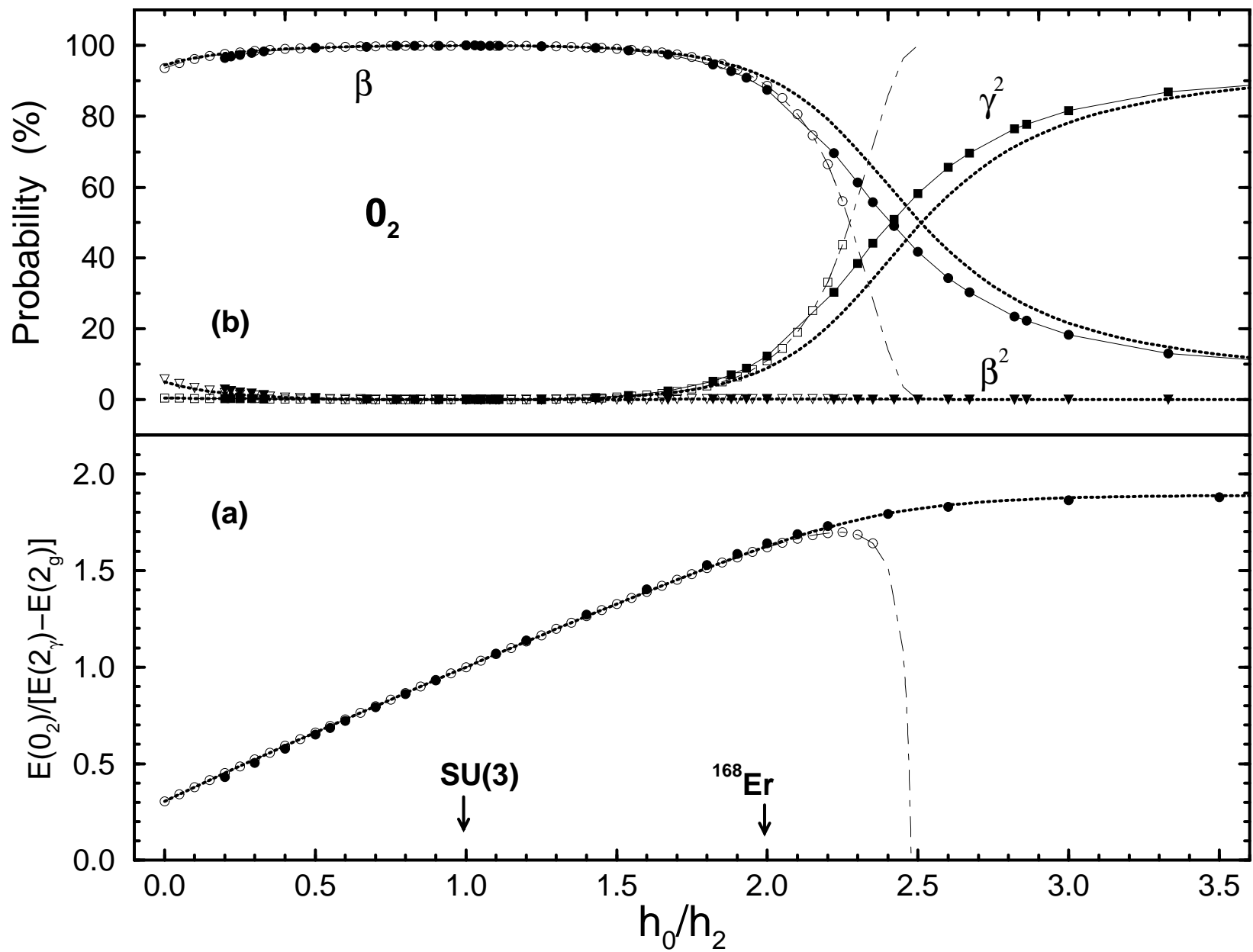
\end{document}